\documentstyle[11pt,IAU207_pasp,twoside]{article}
\def\edcomment#1{\iffalse\marginpar{\raggedright\sl#1\/}\else\relax\fi}
\reversemarginpar

\begin{document}
\title{M31 globular cluster metallicities and ages}

\author{Pauline Barmby}
\affil{Harvard-Smithsonian Center for Astrophysics, 60 Garden St., Mailstop 10, 
Cambridge MA 02138, USA}

\begin{abstract}
Globular cluster ages are more than just lower limits to the age of the universe;
the distribution of ages constraints the timescales for galaxy formation and chemical evolution.
Globular cluster populations with different metallicities have now been detected in many
galaxies, and understanding how these populations formed requires knowing their relative ages.
We examined the relative ages of the two M31 globular cluster populations using their color and 
luminosity distributions and found that the metal-rich clusters could be up to 50\% younger than
the metal-poor clusters. While a small delay in the formation of metal-rich clusters might be
imposed by chemical enrichment timescales, a large age gap
demands a more detailed explanation. I outline several possibilities and their promises
and problems.
\end{abstract}

\section{Introduction}

Studies of the M31 globular cluster system are thoroughly
reviewed by Jablonka (these proceedings). One of the most important discoveries about the
M31 GCS is that, like that of the Milky Way, it contains two populations of globular clusters
with different metallicities, kinematics, and spatial distributions.
Many other galaxies' GCSs also show this bimodality, and it has been 
seized upon as an important key to GCS and galaxy history.
An important question is whether the two populations
of clusters have different ages. 

Measuring ages for individual M31 GCs by fitting isochrones to their 
CMDs is not possible with existing technology, although it is part of
the design reference mission for NGST (Rich \& Neill 1999). Age estimates
using horizontal branch morphologies, as done by Sarajedini et al.\ (2000) for M33
clusters, is possible, although it relies on the (still controversial) 
assumption that age is the dominant `second parameter'.
We used a new catalog of integrated properties of
M31 GCs and GC candidates (Barmby et al.\ 2000) to investigate the 
ages of the metal-poor and metal-rich M31 GCs. An important feature of this catalog
is its use to determine the extinction of individual clusters, and thus
their intrinsic colors and magnitudes. 

\section{Cluster colors and SSP models}
	
The original idea of this work (Barmby \& Huchra 2000) was to compare the 
intrinsic colors of M31 and Milky Way globular clusters to the predictions
of simple stellar population models. This is important not only for eliciting
the properties of M31 GCs, but also for testing the models. GCs are among the
simplest stellar populations known, and models which cannot reproduce their
integrated colors will presumably have problems when used to study composite
populations like galaxies.

The intrinsic colors of M31 globular clusters are taken from 
our catalog (Barmby et al.\ 2000); colors for Milky Way globulars
are from the Harris (1996) catalog, augmented by infrared colors
published in Brodie \& Huchra (1990).
For this comparison, we used only clusters with measured spectroscopic
metallicities and reddening $E(B-V)<0.5$, since
uncertainty in the reddening, and hence the intrinsic colors, increases
for large values of $E(B-V)$. 

We compared the cluster colors to those of simple stellar 
populations from three sets of models:
those of Worthey, Bruzual \& Charlot (both the Worthey and B\&C models are 
the versions 
reported in Leitherer et al.\ 1996), and Kurth, Fritze-von Alvensleben, 
\& Fricke (1999). We used all the available model metallicities,
and model ages of 8, 12, and 16 Gyr. Although model colors are tabulated 
in smaller age increments (typically 1 Gyr), we decided that
it was more reasonable to use only a few typical ages
than attempt to derive precise cluster ages from integrated colors.

The models generally fit the data very well, with offsets
in the bluer colors likely due to model problems. However, the
best-fit models for the metal-rich clusters have younger
ages (8 or 12 Gyr) than those for the metal-poor clusters (16 Gyr).
Unless there is some systematic problem with metallicity in
our data and/or the models, this means that the metal-rich clusters
are younger than the metal-poor clusters in both M31 and the Milky Way.

\section{GCLF variation}

Many factors influence the total luminosity of a globular cluster, including
mass, IMF, age, and metallicity. It is thus somewhat surprising that the
distribution of integrated GC luminosities, the globular cluster luminosity function
or GCLF, is so similar from galaxy to galaxy. 
In Barmby, Huchra, \& Brodie (2001), we computed the GCLF for various
sub-samples of the M31 GCS. We used the intrinsic $V_0$ magnitudes for
clusters with $V<18$ and careful corrections for incompleteness,
and computed the GCLF parameters using the maximum likelihood method
described in Secker \& Harris (1993). 

The peak of the GCLF was brighter for clusters near the center of M31 
than for more distant clusters. The GCLF difference found was comparable
to theoretical predictions of cluster destruction effects 
(e.g., Ostriker \& Gnedin 1997). We also found a difference between
the GCLFs of metal-rich and metal-poor clusters. The metal-rich clusters
are brighter, not fainter as would be expected from the effects
of metallicity on $M/L_V$ (Ashman, Conti, \& Zepf 1995).
One way for the metal-rich clusters to be brighter than the metal-poor
clusters is for them to be significantly younger. Comparisons with 
the population synthesis models mentioned above show that the luminosity
difference we measured could be caused by a difference in age of about 55\%.

\section{Summary}

Integrated colors and luminosities of M31 globular clusters provide
photometric evidence that at least some of the metal-rich GCs in M31
are younger than the bulk of the population. Such an effect has
previously been suggested to explain the ratio of dwarf to giant light 
found in the spectra of metal-rich M31 clusters by Tripicco (1989).
The age differences our results imply are large --- up to 8 Gyr. 
Chemical evolution models generally do not
require this much time to enrich protocluster gas, so we need some
other reason for the large age gap between populations. One possibility 
is that the metal-rich clusters in M31 formed in a merger (as in the
picture of Ashman \& Zepf 1992), but any such merger would have to be
gentle enough not to destroy the M31 disk. Another possibility is that 
the metal-poor clusters were formed before the galaxy itself, and the
metal-rich clusters formed with the galaxy. If this was the case, one
might expect to see similar age differences in the GC populations
of other galaxies. Integrated broad-band magnitudes and colors are clearly not
the best way to age-date globular clusters, although they provide 
tantalizing hints in the case of the M31 GCs. More precise
methods are needed for age-dating extragalactic globular clusters.

\acknowledgements
I thank the conference organizers for a stimulating meeting
and for the financial assistance which enabled me to attend, and
the Smithsonian Institution for financial support.

\end{document}